\documentclass{emulateapj}

\slugcomment{To appear in Astrophysical Journal Letters}

\shorttitle{Discovery of Two Dust Pillars near the Galactic Plane}
\shortauthors{\'Ubeda \& Pellerin}

\begin{document}


\title{Discovery of Two Dust Pillars near the Galactic Plane}


\author{Leonardo \'Ubeda}
\affil{D\'epartement de physique, de g\'enie physique et 
d'optique, Universit\'e Laval, and Observatoire du  Mont M\'egantic, Qu\'ebec, G1K 7P4, Canada.} 
\and
\author{Anne Pellerin}
\affil{Space Telescope Science Institute, 3700 San Martin Drive, Baltimore, MD 21218, USA}


\begin{abstract}
We report the discovery of two dust pillars using GLIMPSE archival  images
obtained with the Infrared Array Camera on board  the {\it Spitzer Space Telescope.}
They are located close to the  Galactic molecular cloud  GRSMC45.453+0.060
and they   appear to be aligned  with the ionizing region 
associated with GRSMC45.478+0.131.
Our three colour mosaics show that these stellar incubators   present  different  morphologies
as seen from planet Earth.
One of them shows the  unquestionable existence of young stellar objects
in its head, whose  influence on the original cocoon  is evident, while the other
presents a well defined bright-rimmed ionizing front. 
We argue that  second-generation star formation  has been triggered in these protuberances
by the action of massive stars present in the nearby H~{\scriptsize II} regions.

\end{abstract}



\section{Introduction}

The Milky Way presents a region at approximately 5 kpc from
the Sun that dominates its star-forming structure.
This region is called the Galactic Ring and it 
is an enormous  reservoir of material for the formation of new stars and stellar 
 clusters. Most of the Galactic Giant H~{\scriptsize II} regions, far-infrared (far-IR)  luminosity, 
diffuse ionized gas, and supernovae remnants are associated with the ring 
according to \cite{Clemetal88}.

The Galactic Ring  remains largely unexplored. 
 \citet{Jacketal06} have conducted a molecular line survey of the inner Galaxy
called the Boston University-Five College Radio Astronomy Observatory
 Galactic Ring Survey (GRS)\footnote{This publication makes use of molecular line data from the Boston University-FCRAO Galactic Ring Survey (GRS). The GRS is a joint project of Boston University and Five College Radio Astronomy Observatory, funded by the National Science Foundation under grants AST-9800334, AST-0098562, \& AST-0100793.}. 
This survey uses the molecular transition  $^{13}$CO~$J=1~\rightarrow~0$ which
peers more deeply into the interiors
of molecular clouds than the more common  $^{12}$CO.
This data set has provided the tools
to discover infrared dark clouds which represent 
the densest clumps within giant molecular clouds where massive stars may 
eventually form. We are using this information to select  
several regions that present  high intensity in the molecular transition
in order  to perform 
a detailed mid- and near-IR follow-up research. 

We are interested in performing a thorough  description
of the stellar populations in those regions, with emphasis on the massive stellar components.
Massive stars are born in giant molecular clouds and form most likely
 in stellar clusters \citep{LadaLada03}.
They physically and chemically interact with their environment by the ionization
due to  ultraviolet (UV)  light, by stellar winds, and by supernova explosions.
Massive stars spend their youth embedded in the molecular cloud in which 
they form; therefore, young massive stars must be studied at infrared or longer wavelengths.

It has been found that massive young clusters 
$(M > 10^4 \, M_{\odot} $ and age $< 20$ Myr)
usually trigger a second-generation
of star formation, and frequently the most luminous second-generation 
sources are located inside  the heads of dust pillars oriented towards
the central cluster. 
Giant dust pillars are often found at the edges of H~{\scriptsize II}  regions. 
The dust pillars in M16 (the Eagle Nebula, \citealt{Hestetal96}) are a 
well known example.   \cite{Walbetal02b}
provides a review of the some  interesting dust pillars in our  Galaxy
 including
the Horsehead, the Cone,
the Eagle, the Carina Finger, and one in NGC 3603. Dust pillars have also been found 
in starburst regions 
 in the Magellanic Clouds (30 Doradus and N11).

In this Letter, we present the discovery of  two  dust pillars
located close to the conspicuous  molecular cloud  GRSMC45.453+0.060  (using the notation of 
\citealt{Simoetal01} to name the  Galactic Ring 
Survey Molecular Clouds.) 
We briefly discuss
their morphology from the observation  of
 three-colour mosaics constructed using {\it Spitzer  Space Telescope} images.

\section{Observations}

\subsection{Galactic Ring Survey}
 \citet{Jacketal06}  have conducted a molecular line survey of the inner Galaxy
using the SEQUOIA multipixel array on the
Five College Radio Astronomy Observatory 14--m telescope.
This  Galactic Ring Survey (GRS) uses the molecular transition 
 $^{13}$CO~$J=1~\rightarrow~0$,
and   covers   a Galactic 
 longitude range of $l=18\fdg0 -55\fdg7$ and a 
latitude range of $|b|<1^\circ$, a total of 75.4 deg$^2$.

We have used the GRS
to identify  interesting regions  where we would  perform mid- and near-IR studies, with 
particular interest in the massive star populations.
The 5 kpc ring dominates both the molecular interstellar medium and 
the star-formation activity in the Milky Way. Therefore, it  plays a crucial role in the dynamics, 
structure, and evolution of our Galaxy.

We retrieved the FITS cubes in the range 
$l~=~43\fdg0-47\fdg0$, $b~=~0\fdg0$ from their survey and we built a 
colour contour plot shown in Figure~1. 
The image shows the  GRS $^{13}$CO  intensity integrated over all velocities ($V_{LSR} = -5$ to 85~km~s$^{-1}$). 
The color scale ranges from 0 to 30 K km s$^{-1}$.   \citet{Jacketal06}   gives the details of these observations. 
Four molecular clouds are clearly seen and we labeled them in the bottom 
image. These clouds have  been studied by \citet{Simoetal01} and \citet{Kraeetal03} using the 
GRS and the  {\it Midcourse Space  Experiment } data.

\subsection{Spitzer observations }
The Galactic Legacy Infrared Mid-Plane Survey 
Extraordinaire (GLIMPSE;  PI: Ed Churchwell, see \citealt{Benjetal03} for a description 
of the project),  was conducted  using the
 Infrared Array Camera (IRAC; \citealt{Fazietal04})  onboard  the {\it Spitzer 
Space Telescope} \citep{Wernetal04}. This survey covered  approximately 220 deg$^2$
 of the Galactic plane, covering a latitude
range of $b= \pm 1^\circ$, and a longitude range of $|l|=10\fdg0-65\fdg0$.
IRAC has four bands, centered at approximately 3.6, 4.5, 5.8 and 8.0 $\mu$m
(hereafter [3.6], [4.5], 
[5.8], and [8.0], respectively),
each of which has a field of view of $\sim 5\farcm2 \times 5\farcm2$. All four bands
are observed simultaneously. 
The calibrated data from
the {\it  Spitzer }
Science Center (SSC pipeline ver. S10.5.0)
were processed through the GLIMPSE pipeline reduction 
system. 
Image processing includes masking hot, dead, and missing data 
pixels. Several image 
artifacts (described in \cite{Horaetal04}  and the IRAC Data Handbook) are corrected for in the 
GLIMPSE pipeline:  they remove artifacts such as stray light in all bands, they correct for muxbleed  
(in [3.6] and [4.5] bands), and for banding (in [5.8] and [8.0] bands). 
Cosmic rays are also removed.

The individual frames were mosaicked using 
{\sc\footnotesize MONTAGE}\footnote{See http://montage.ipac.caltech.edu. This research made use of Montage, funded by the National Aeronautics and Space Administration's Earth Science Technology Office, Computation Technologies Project, under Cooperative Agreement Number NCC5-626 between NASA and the California Institute of Technology. Montage is maintained by the NASA/IPAC Infrared Science Archive.},  to produce an 
image of the entire field at each band.
The mosaic images 
conserve surface brightness of the original images. 
The angular size of each resulting tile is  $1\fdg1 \times  0\fdg8$ or $6640 \times 4840$ pixels$^2$.
The pixel size of the final data product  is $0\farcs6$.
For this work, we used  the archival images from the GLIMPSE survey  which are centered
around $l=45\fdg50$, $b = 0\fdg0$ and which are provided by the {\it Spitzer}  Science Center in their
data release of April 2007.


\subsection{2MASS images }
The Two Micron All Sky Survey\footnote{This publication makes use of data products from the Two Micron All Sky Survey, which is a joint project of the University of Massachusetts and the Infrared Processing and Analysis Center/California Institute of Technology, funded by the National Aeronautics and Space Administration and the National Science Foundation. } (2MASS) \citep{Skruetal06} 
 has uniformly scanned the 
entire sky in three near-infrared bands to detect and characterize point 
sources brighter than about 1~mJy in each band, with signal-to-noise ratio
 greater than 10, using a pixel size of $2\farcs0.$ 
This project used two 1.3--m telescopes (one in each hemisphere) equipped with a three-channel camera, 
each channel consisting of a 256 $ \times$ 256 pixels$^2$  
array of HgCdTe detectors, capable of 
observing the sky simultaneously in bands 
 $J$ (1.25~$\mu$m), $H$ (1.65~$\mu$m), and $K_S$ (2.17~$\mu$m).
We retrieved  $J, H$, and $K_S$ archival 2MASS images 
 that cover a 30 arcminute region around $l=45\fdg50$, $b = 0\fdg0$.
 We also downloaded the catalogue of stars from the Point Source Catalog (PSC) 
  that contains a total 
 of 79528 objects found in the same region.

\section{Mid-infrared mosaics}
The individual FITS  frames from GLIMPSE were mosaicked using 
{\sc\footnotesize MONTAGE}  and our own IDL codes  to produce a three-colour 
image of the entire field around GRSMC45.453+0.060.
We combined   the    images obtained in filters
[3.6],  [5.8],  and [8.0] by assigning  blue to the shortest wavelength  band, green
to the intermediate band, and red to the longest wavelength band.
This particular color combination was chosen to highlight particular properties of the 
images.
Figure~1 [Bottom] shows the three-colour  mosaic 
where we have labeled four spectacular  molecular clouds: GRSMC45.073+0.129, GRSMC45.122+0.132,
GRSMC45.453+0.060, and GRSMC45.478+0.131. Each of them is associated with an 
{\it Infrared Astronomical Satellite (IRAS)} point source and with thermal radio continuum emission 
which is indicative of H~{\scriptsize II}  regions \citep{Kraeetal03}. 
A comparison with  the  GRS $^{13}$CO $J=1 \rightarrow 0$ map shows that all four molecular cores present
bright $^{13}$CO emission as well as  extended mid-IR emission.
Note the bridge of $^{13}$CO emission connecting the two groups of molecular clouds.

A closer inspection of the structure close to GRSMC45.453+0.060
reveals the existence of two  striking  dust pillars. Figure~2 shows a magnified version 
of this region in the Galaxy.  

The IRAC bands show bright, diffuse, and very highly structured emission, which is 
produced primarily by polycyclic aromatic hydrocarbon (PAH) features and Br$\alpha$ line emission
\citep{Drai03}.


The reddish shades in Figure 2 indicate the distribution of warm dust emitting primarily in the 6.2$\mu$m PAH 
feature in the [5.8] band and the 7.7 and 
8.6$\mu$m PAH features in the [8.0] band. 
Main-sequence stars with photospheres hotter than
1000~K appear blue in IRAC colors, except for the most reddened 
ones, and are brightest at [3.6] and [4.5]   \citep{Churetal04}.
We observe that the diffuse emission 
systematically and sharply increases from [3.6] to [8.0] and that it is extremely faint 
or inexistent in the 2MASS  $J, H$, and $K_S$   images.

\section{Morphology of the dust pillars}
Our three-colour mosaic, built using mid-IR GLIMPSE images in IRAC [3.6], [5.8],  and [8.0] bands shows
the existence of a couple of dust pillars located to the northeast of  GRSMC45.453+0.060. Using the 
approximate Galactic coordinates
of their heads, we name them DP45.542-0.006 and DP45.543-0.029. Table~1 provides
their Galactic coordinates.

We may argue   that these pillars  appear to be aligned  with the ionizing region GRSMC 45.478+0.131.
We can also speculate that   DP45.543-0.029 is located in front of DP45.542-0.006, and is more
heavily obscured. The northern rim of  DP45.542-0.006
is very well defined and we can trace the silhouette of  DP45.543-0.029 on the structure
of DP45.542-0.006.
These protrusions obviously began as pre-existing
density enhancements in the original  surrounding molecular
cloud, and they evolved into the forms seen under the combined effect
of photoionization and flows coming from the nearby H~{\scriptsize II} regions. 
These pillars
are the result of  the interaction of massive stars with their environment,
and they are most likely places of star formation. 


For our research we assume a distance to the 
molecular cloud  GRSMC45.453+0.060 of 6.0 kpc. This distance was determined kinematically
by \citet{Simoetal01}.
A simple  measurement of the length and width of the pillars is given in Table~1.
The definition of these measurements is of course subjective  because the base of the pillars
merges into larger dust structures. We want to emphasize that  we measure projected
quantities, since the inclination angles are unknown.
Our estimated values are in agreement with those that   \citet{Walbetal02b}   calculated for other
dust pillars in the Galaxy.

The heads of the two pillars  that we discovered are dramatically different as seen
from our point of view. 
Both of them show the presence 
of  young stellar objects, but while the stars in the head of  DP45.543-0.029 appear to  remain
enclosed in the dusty cocoon, those of  DP45.542-0.006 have already started to 
 disrupt  the  interstellar medium. This is a perfect indicator of
ongoing stellar evolution, and it may suggest  that the stars in the latter pillar are  older.
However, it is possible that due to the geometry of the region, the dust may be
concealing the disruption of the ISM produced by the new born stars in   DP45.543-0.029.
From the 2MASS point source catalogue of this region, and at 
a resolution of  $2\farcs0$ per pixel we find three objects at the tip of  DP45.542-0.006, and 
two in DP45.543-0.029.

We may speculate that the stars in 
these   pillars represent a second  generation of star formation 
in this region. Studies of bright-rimmed clouds containing infrared sources have
been taken as evidence of star formation induced by radiatively driven implosion 
 \citep{Walbetal02b,Hestetal96,SugiOgur94}.
 Triggered second-generation star formation in the vicinity of O-type stars
and clusters is ubiquitous, and often the second generation is associated with dust pillars.
These processes may play an important role
in the self propagation of star formation in the Galaxy.

\section{Results and future work}
Recent  Galactic surveys in radio and the infrared are providing us with exciting new tools 
to study nearby star formation and the interaction of massive stars with their environment. 
In this research we have used the Galactic Legacy Infrared Mid-Plane Survey 
Extraordinaire (GLIMPSE) conducted using  the {\it Spitzer 
Space Telescope} and the  Galactic Ring Survey (GRS) performed at the 
Five College Radio Astronomy Observatory. 
GRS has the ability to identify and probe active star-forming gas, and GLIMPSE
delineates the structure of molecular clouds and their dust composition. 

We discovered two remarkable dust pillars  near the molecular cloud 
GRSMC45.453+0.060. 
Even though they seem close to each other (in projection), one of them shows a more evolved structure that
the other, from our point of view.

We intend to  continue our research of this whole star forming region of the Galaxy.
We will perform near infrared studies
to gather information on their  stellar populations by means of   photometry and spectroscopy.
A future paper will investigate the truth of the 
speculations proposed in this Letter  in more detail, but new spectroscopic and photometric 
observations of this region 
are strongly encouraged in order to characterize the nature of these clouds
and the stellar population associated with them.  
The space-mass-age distributions of the
objects in these regions will tell us a great deal about the relationships between 
higher- and lower-mass star formation.
In particular we are interested in the  distribution of stellar ages  in order to  corroborate our hypothesis
of the two-stage star formation process. 

The early results of our  research  
indicate  the  importance of large 
field-of-view IR observations like the ones provided by the GLIMPSE survey. 
We are confident that our detailed  studies  of
star forming regions near the Galactic plane   will play a crucial role in the 
overall understanding of the Milky Way.

\acknowledgments

Leonardo \'Ubeda acknowledges funding from 
the  Fonds qu\'eb\'ecois de la recherche sur la nature et les technologies.
This work was supported by HST grant HST-AR-10968.02-A.






\begin{deluxetable}{lccccc}
\tabletypesize{\scriptsize}
\tablecaption{Coordinates and morphological parameters of the dust pillars}
\tablewidth{0pt}
\tablehead{  \colhead{Pillar} & \colhead{$l$} &\colhead{$b$} &\colhead{Length} & \colhead{Width} & \colhead{L/W}  \\
                        \colhead{} & \colhead{ 45\degr 32\arcmin+} &\colhead{ $-$00\degr+} &\colhead{[pc]} & \colhead{[pc]} & \colhead{}   }

\startdata
DP45.542-0.006 &$32\arcsec$& 00\arcmin 21\arcsec& 2.3 & 1.1 & 2.1    \\
DP45.543-0.029 &$34\arcsec$& 01\arcmin45\arcsec& 1.7 & 0.9 & 1.9    \\

\enddata


\end{deluxetable}

\begin{figure}
 \centering
\includegraphics[width=140mm]{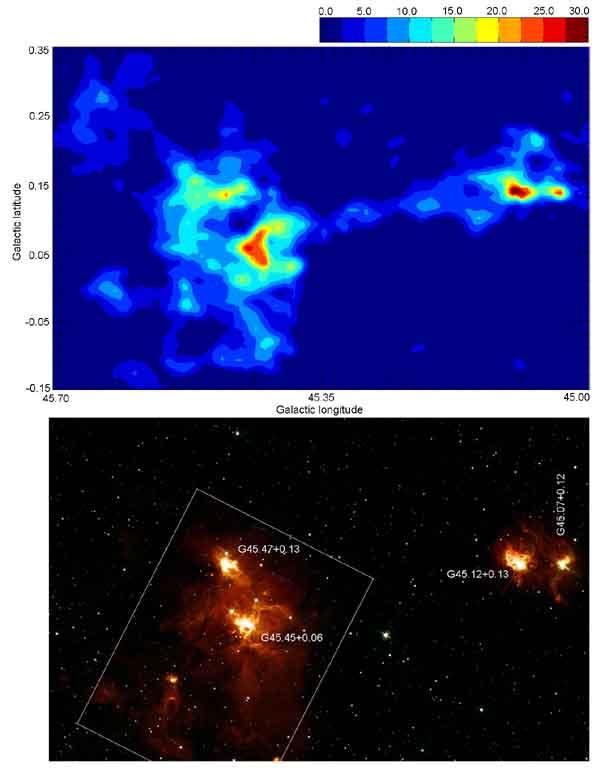}
\caption{  [Top] Color-scale representation of the integrated intensity of the
$^{13}$CO $J=1 \rightarrow 0$ emission from the Galactic Ring Survey, integrated from 
$V_{LSR} = -5$ to 85~km~s$^{-1}$. The colour scheme ranges from 0 to 30~K~km~s$^{-1}$.
The orientation  is given in the  Galactic coordinate system. 
[Bottom]  Three colour mosaic built with  IRAC images. We combined  images obtained with  filters 
3.6$\mu$m (blue channel), 5.8$\mu$m (green channel),  and 8.0 $\mu$m (red channel). 
We have labeled four  prominent  molecular clouds, and the rectangle shows the enlarged area
depicted in Figure~2.   }
\end{figure}

 \begin{figure}
 \centering
\includegraphics[width=160mm]{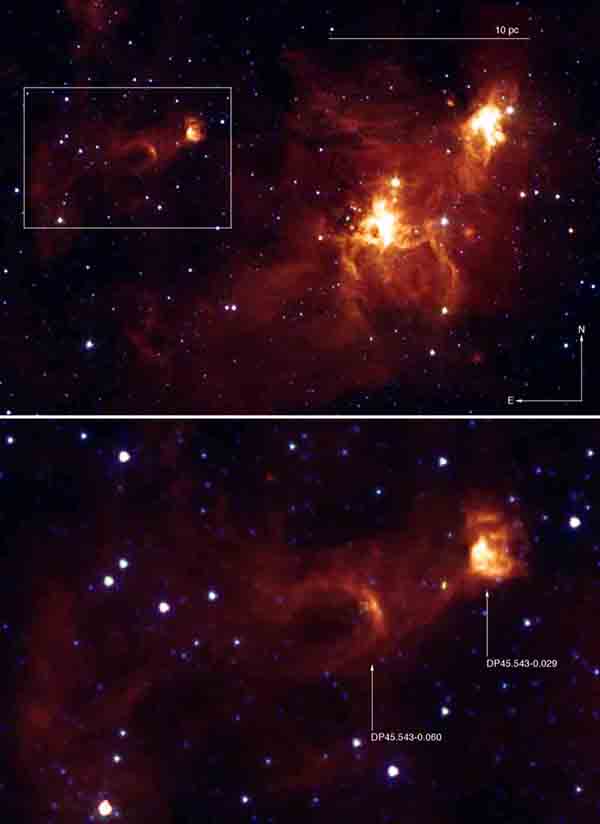}
\caption{ [Top] This three colour mosaic  shows the region in Figure 1 marked with the rectangle. 
We have rotated this image so that North is up and East is left. [Bottom] A zoomed version of the 
mosaic showing in detail the structure of the two dust pillars.  Note the different morphology 
of the two heads as we see  them from planet Earth.}
\end{figure}









\end{document}